\documentclass[%
 reprint,
 superscriptaddress,
 amsmath,amssymb,
 aps, prapplied,
 floatfix
]{revtex4-2}

\usepackage[utf8]{inputenc}
\usepackage[T1]{fontenc}
\usepackage{textcomp}
\usepackage{graphicx}
\usepackage{dcolumn}
\usepackage{bm}
\usepackage{hyperref}
\usepackage{xcolor}

\begin{document}

\author{Taewoong Yoon}
\affiliation{Department of Physics and Astronomy, Seoul National University,
Seoul 08826, Korea}
\affiliation{Institute of Applied Physics, Seoul National University, Seoul
08826, Korea}

\author{Sangwon Oh}
\affiliation{Department of Physics and Department of Energy Systems Research,
Ajou University, Suwon 16499, Korea}

\author{Dohun Kim}
\affiliation{Department of Physics and Astronomy, Seoul National University,
Seoul 08826, Korea}
\affiliation{Institute of Applied Physics, Seoul National University, Seoul
08826, Korea}

\author{Junghyun Lee}
\email{Contact author: jh\_lee@kist.re.kr}
\affiliation{Center for Quantum Technology, Korea Institute of Science and
Technology, Seoul 02792, Korea}

\author{Hyunyong Choi}
\email{Contact author: hy.choi@snu.ac.kr}
\affiliation{Department of Physics and Astronomy, Seoul National University,
Seoul 08826, Korea}
\affiliation{Institute of Applied Physics, Seoul National University, Seoul
08826, Korea}

\date{\today}

\title{Mesoscopic Spin Coherence in a Disordered Dark Electron Spin Ensemble}

\begin{abstract}
Harnessing dipolar spin environments as controllable quantum resources is a
central challenge in solid-state quantum technologies. Here, we report the
observation of a coherent mesoscopic spin state in a disordered ensemble of
substitutional nitrogen (P1) centers in diamond. An iterative Hartmann-Hahn
protocol transfers polarization from dense nitrogen-vacancy (NV) centers to a P1
ensemble, yielding a 740-fold enhancement over room-temperature thermal
equilibrium as revealed by differential readout. The resulting mesoscopic P1
spin ensemble exhibits collective Rabi oscillations and long-lived spin-lock and
Hahn-echo coherences. We identify a crossover in the saturation polarization
arising from the competition between coherent driving and local disorder,
providing an estimate of the local disorder scale. These results establish a
foundation for utilizing dark electron spin ensembles as robust resources for
quantum sensing and quantum many-body simulation.
\end{abstract}

\maketitle

\section{Introduction}

Solid-state spin systems have emerged as a compelling platform for quantum
information science, ranging from nanoscale magnetometry to probing many-body
dynamics~\cite{degen_quantum_2017,wolfowicz_quantum_2021,%
chatterjee_semiconductor_2021,awschalom_quantum_2018,rovny_nanoscale_2024}.
Although surrounding spin environments are typically regarded as a primary
source of decoherence~\cite{bauch_decoherence_2020,park_decoherence_2022}, their
large Hilbert space can be exploited as a powerful quantum resource. This
potential motivates the use of proximal nuclear and electronic spins as
long-lived quantum memories and
registers~\cite{bradley_ten-qubit_2019,bourassa_entanglement_2020,%
degen_entanglement_2021,ungar_control_2024}, the realization of environment- and
logic-assisted sensing protocols in which ancillary spins enhance
sensitivity~\cite{jiang_repetitive_2009,arunkumar_quantum_2023,%
cooper_environment-assisted_2019}, and the exploitation of dense spin ensembles
as analog quantum simulators of emergent
phenomena~\cite{pagliero_optically_2020,zu_emergent_2021,davis_probing_2023}.

Among spin environments, substitutional nitrogen (P1) centers constitute the
dominant paramagnetic defects in diamond, typically outnumbering
nitrogen-vacancy (NV) centers owing to the limited conversion efficiency during
synthesis~\cite{mindarava_efficient_2020,osterkamp_benchmark_2020,%
luo_creation_2022}. In contrast to well-established NV spin qubits with robust
optical initialization and readout~\cite{doherty_nitrogen-vacancy_2013}, the
more abundant P1 centers form an optically dark electronic spin bath that limits
NV coherence~\cite{park_decoherence_2022, bauch_decoherence_2020}. Exploiting
these dark spins as a quantum resource requires polarization transfer from NV
centers, commonly achieved via resonant spin flip-flop interactions at specific
external magnetic fields~\cite{hanson_polarization_2006, loretz_optical_2017,
zu_emergent_2021}. Alternatively, spin-locking (SL) under the Hartmann-Hahn (HH)
condition offers a more versatile pathway for polarization exchange in the
rotating frame~\cite{hartmann_nuclear_1962,
belthangady_dressed-state_2013,laraoui_approach_2013}. This method enables
control over both the coupling strength and the effective
disorder~\cite{kucsko_critical_2018,lee_dressed-state_2023}. However, previous
approaches often rely on isolated NV centers to polarize local
environments~\cite{laraoui_approach_2013, knowles_demonstration_2016}, where
polarization rapidly dissipates via spin diffusion into the bulk thermal
reservoir~\cite{loretz_optical_2017, ranjan_probing_2013, prisco_scaling_2021}.
As a result, robust mesoscopic polarization in disordered P1 ensembles has
remained unrealized.

In this work, we demonstrate the generation and control of a coherent mesoscopic
spin state in a disordered ensemble of dark P1 centers in diamond. To prepare
and interrogate this state, we employ an iterative HH protocol combined with a
polarization-direction-differential readout scheme. Leveraging a dense NV
network, we achieve a 740-fold enhancement of P1 ensemble polarization over the
room-temperature thermal limit. The rapid saturation observed within a few
transfer cycles indicates that additional relaxation channels during NV
initialization play a critical role in the P1 polarization process. We observe
collective Rabi oscillations and characterize robust SL relaxation and Hahn-echo
coherence times. Moreover, we identify a crossover in the saturation
polarization where coherent driving overcomes local disorder to enable efficient
polarization exchange. Together, these results show that a disordered dark-spin
ensemble can be prepared in a robust polarized state and coherently controlled,
and identify the conditions required for efficient polarization transfer.

\begin{figure}
\includegraphics{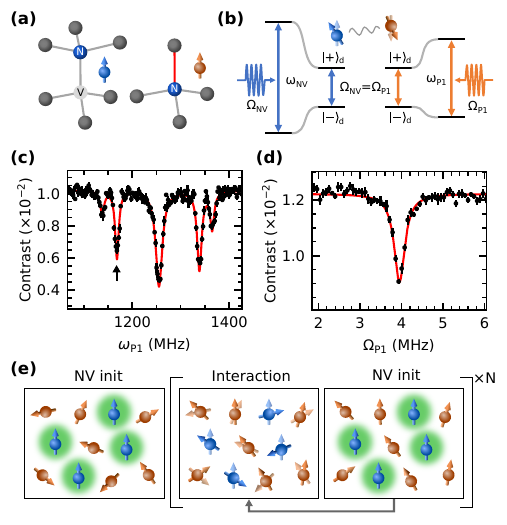}
\caption{NV-P1 hybrid electron spin system and polarization transfer protocol.
(a) Atomic structures of the NV (left) and P1 (right) centers. Spheres represent
carbon atoms (gray), nitrogen atoms (blue), and vacancies (light gray). The P1
JT axis is highlighted in red. Blue and orange arrows denote NV and P1 electron
spins, respectively. (b) HH condition. Two spin species with distinct resonance
frequencies $\omega_\text{NV}$ and $\omega_\text{P1}$ are resonantly driven at
Rabi frequencies $\Omega_\text{NV}$ and $\Omega_\text{P1}$, respectively. In the
dressed-state basis $\{|+\rangle_d, |-\rangle_d\}$, the energy splittings are
matched when $\Omega_\text{NV} = \Omega_\text{P1}$. (c) P1 electron spin
resonance spectrum measured via DEER. The hyperfine interaction and JT
distortion result in five visible spectral subgroups with population fractions
of $\{1,3,4,3,1\}/12$. We address a P1 subgroup with a population fraction of
3/12 (indicated by an arrow) for polarization transfer measurements. The solid
red line represents a multi-Lorentzian fit. (d) Measurement of the HH resonance
condition. NV SL contrast is plotted as a function of the P1 Rabi frequency
$\Omega_\text{P1}$, with $\Omega_\text{NV}=3.95$~MHz. The solid line represents
a Lorentzian fit. (e) Iterative polarization transfer protocol. The sequence
comprises optical NV initialization followed by interaction with the P1 ensemble
under the HH condition. This cycle is repeated to accumulate polarization in the
P1 ensemble. Error bars represent s.e.m. \label{fig:1}}
\end{figure}

\section{Experimental system}

Our experiments are conducted on a hybrid electron spin ensemble embedded in an
isotopically purified diamond substrate at room temperature
[Fig.~\ref{fig:1}(a)]. The sample contains high concentrations of P1 centers
($[\text{P1}] = 3.1 \pm 0.1$~ppm) and NV centers ($[\text{NV}] = 1.2 \pm
0.2$~ppm) (Appendix~\ref{sec:concentration}), corresponding to $\sim10^4$ spins
within the probed volume. Each P1 center hosts an optically dark electron spin
($S=1/2$) subject to strong hyperfine coupling to its host $^{14}\text{N}$
nucleus ($I=1$) and Jahn-Teller (JT)
distortion~\cite{de_lange_controlling_2012,degen_entanglement_2021}. We apply a
bias magnetic field of $B_z = 446$~G along the NV symmetry axis to lift the
spin-state degeneracy, allowing us to treat the $\{|0\rangle, |-1\rangle\}$
subspace of the spin-1 NV ground state as an effective spin-1/2 system. 

Due to the substantial Zeeman energy mismatch between the two electron spin
species, the dipolar Hamiltonian is reduced to a secular Ising interaction,
$H_{\text{int}} = \sum_{i<j} J_{ij} S_i^z S_j^z$, where $S_i$ denotes the spin
operator of the $i$-th spin and $J_{ij}$ is the dipolar coupling strength. To
bridge the energy mismatch, we employ SL, where resonant driving defines a
dressed-state energy splitting equal to the Rabi frequency
$\Omega$~[Fig.~\ref{fig:1}(b)]. Under driving along the $x$-axis, which defines
the effective quantization axis in the rotating frame, the longitudinal Ising
coupling is converted into spin flip-flop interactions, as the original $S^z$
operators become transverse in the dressed-state basis. When the Rabi
frequencies satisfy the HH condition ($\Omega_{\text{NV}} = \Omega_{\text{P1}} =
\Omega$), the NV-P1 system undergoes resonant spin
exchange~\cite{belthangady_dressed-state_2013,laraoui_approach_2013}. In the
strong driving limit ($\Omega \gg W$), the effective disorder in the dressed
frame is reduced to $W_{\text{eff}} \approx W^2 / (2\Omega)$, where $W$ denotes
the characteristic disorder
scale~\cite{kucsko_critical_2018,rezai_probing_2025}.

We characterize the energy spectrum of the dark P1 ensemble using double
electron-electron resonance (DEER), and the HH resonance using double
spin-locking (DSL)~\cite{de_lange_controlling_2012,
belthangady_dressed-state_2013}. The DEER spectrum exhibits five distinct
subgroups with population fractions of $\{1,3,4,3,1\}/12$, arising from the
three $^{14}\text{N}$ nuclear spin states and the four JT distortion axes
[Fig.~\ref{fig:1}(c)]. For the subsequent experiments, we focus on the subgroup
with a $3/12$ population fraction. The HH resonance is experimentally verified
by a pronounced Lorentzian dip in the DSL contrast when the Rabi frequencies are
matched [Fig.~\ref{fig:1}(d)]. Based on these results, we implement an iterative
protocol comprising repetitive cycles of optical NV initialization and HH
interaction to accumulate polarization in the P1 bath [Fig.~\ref{fig:1}(e)].
Here, NV centers act as both optically active polarization sources and sensors
to read out the collective state of the P1 ensemble. 

\section{Experimental setup}
\label{sec:experimental}

Experiments are performed at room temperature on a chemical vapor deposition
(CVD) diamond substrate (Element Six) containing a 40~\textmu{}m thick, 99.9\%
$^{12}\text{C}$-enriched epitaxial layer. Isotopic purification suppresses
$^{13}\text{C}$ nuclear spin bath contributions, such that the electron spin
dynamics are predominantly governed by dipolar interactions between NV centers
and P1 centers.

Optical initialization and readout are conducted using a home-built confocal
microscope. A 532~nm continuous-wave laser (CNI MLL-III-532-200mW) serves as the
excitation source, with intensity and pulse timing controlled by an
acousto-optic modulator (Crystal Technology 3200-147) in a double-pass
configuration achieving an extinction ratio $>50$~dB. The laser power is set to
10~\textmu{}W for initial NV polarization and fluorescence readout, and elevated
to 50~\textmu{}W during polarization transfer cycles for faster NV
re-initialization. The beam is focused onto the sample via an oil-immersion
objective (NA 1.42, Olympus PLAPON60XO). Photoluminescence is collected through
the same objective, spectrally filtered (650--792~nm), and detected by a
single-photon counting module (Excelitas SPCM-ARQH-14-FC).

The probed volume is estimated from the diffraction-limited excitation focus.
For $\lambda = 532$~nm, $\mathrm{NA} = 1.42$, and the diamond refractive index
$n = 2.42$, the full widths at half maximum of the focal intensity are $d_{xy}
\simeq 0.51\lambda/\mathrm{NA} \approx 0.19$~\textmu{}m laterally and $d_z
\simeq 0.88\lambda/(n - \sqrt{n^2 - \mathrm{NA}^2}) \approx 1.0$~\textmu{}m
axially, corresponding to a characteristic volume $V \simeq (\pi/6)\,d_{xy}^2
d_z \approx 0.02$~\textmu{}m$^3$.

Independent addressing of NV and P1 centers is achieved using two microwave
sources. Vector signal generators (SRS SG396 and Anritsu MG3700A) provide
carrier frequencies, modulated by an arbitrary waveform generator (Zurich
Instruments HDAWG) to synthesize amplitude- and phase-modulated pulses. The
signals are combined (Mini-Circuits ZN2PD2-63-S+), amplified by 45~dB
(Mini-Circuits ZHL-16W-43-S+), and delivered to the diamond through a gold
coplanar waveguide lithographically patterned on a glass coverslip.

A static magnetic field of 446~G is applied and aligned along the [111] crystal
axis using three cylindrical samarium-cobalt magnets mounted on motorized linear
stages (Zaber X-LRM-100A-DE51). All experimental sequences, laser modulation,
and gated photon counting are synchronized by the HDAWG operating at 2.4~GS/s.

\section{Effective Hamiltonians for the NV-P1 system}
\label{sec:hamiltonian}

The magnetic dipole-dipole interaction between electron spins $i$ and $j$
separated by distance $r_{ij}$ is
\begin{equation}
H_{\text{dip}}^{ij} = -\frac{J_0}{r_{ij}^3} \left[3(\mathbf{S}_i \cdot
\hat{\mathbf{n}})(\mathbf{S}_j \cdot \hat{\mathbf{n}}) - \mathbf{S}_i \cdot
\mathbf{S}_j \right],
\end{equation}
where $J_0 = \mu_0 \gamma_e^2 \hbar/(4\pi) \approx 52$~MHz$\cdot$nm$^3$ is the
dipolar coupling constant and $\hat{\mathbf{n}} = \mathbf{r}_{ij}/r_{ij}$ is the
unit inter-spin vector. We derive the effective Hamiltonians in two successive
steps: the secular approximation in the rotating frame, followed by the
dressed-state transformation under continuous driving.

\subsection{Dipolar interactions in the rotating frame}

Transforming to the rotating frame via $U(t) = \exp(i\sum_k \omega_k S_k^z t)$,
the spin operators evolve as $S^z \to S^z$ and $S^\pm \to S^\pm e^{\pm i\omega
t}$, where $S^\pm = S^x \pm iS^y$. Expressing the dipolar Hamiltonian in terms
of these ladder operators and applying the secular approximation yields
different effective Hamiltonians depending on the frequency mismatch between the
interacting spins.

For spins sharing the same resonance frequency, such as NV centers along a
common crystallographic axis or P1 centers, the degenerate Larmor frequencies
($\omega_i = \omega_j$) render the flip-flop terms time-independent while the
double-quantum terms oscillate at $2\omega$ and average to zero:
\begin{equation}
\begin{split}
H_{\text{intra}}^{ij} =
-\frac{J_0}{r_{ij}^3}\biggl[&\frac{1-3\cos^2\theta_{ij}}{4}(S_i^+ S_j^- + S_i^-
S_j^+) \\
&+ (3\cos^2\theta_{ij}-1)S_i^z S_j^z\biggr],
\end{split}
\label{eq:intra}
\end{equation}
where $\theta_{ij}$ is the angle between $\mathbf{r}_{ij}$ and the quantization
axis. Defining the effective dipolar coupling $J_{ij} =
J_0(1-3\cos^2\theta_{ij})/r_{ij}^3$, the flip-flop terms mediate
energy-conserving spin exchange with strength $J_{ij}/4$, while the Ising term
$S_i^z S_j^z$ produces configuration-dependent energy shifts.

For NV-P1 pairs, the large frequency mismatch $|\omega_{\text{NV}} -
\omega_{\text{P1}}| \sim 370$~MHz, arising from the NV zero-field splitting
($\approx 2.87$~GHz) and the external magnetic field ($B = 446$~G), suppresses
the flip-flop terms, leaving a purely Ising interaction:
\begin{equation}
H_{\text{inter}}^{ij} = -\frac{J_0}{r_{ij}^3}(3\cos^2\theta_{ij}-1) S_i^z S_j^z.
\label{eq:inter}
\end{equation}
This causes the P1 bath to act as a fluctuating magnetic field for the NV
ensemble, contributing to dephasing but precluding coherent polarization
transfer.

\subsection{Effective interactions in the dressed frame}

In the double spin-locking protocol, continuous microwave drives are applied at
the NV and P1 resonance frequencies, adding
\begin{equation}
H_{\text{drive}} = \Omega_{\text{NV}} S_{\text{NV}}^x + \Omega_{\text{P1}}
S_{\text{P1}}^x
\end{equation}
to the rotating-frame Hamiltonian, where $\Omega_{\text{NV}}$ and
$\Omega_{\text{P1}}$ are the Rabi frequencies. When $\Omega \gg |J_{ij}|$, the
drive defines a new quantization axis along $x$. We introduce tilted-frame
ladder operators $\tilde{S}^\pm = S^y \pm i S^z$, with inverse relations $S^y =
(\tilde{S}^+ + \tilde{S}^-)/2$ and $S^z = (\tilde{S}^+ - \tilde{S}^-)/(2i)$. In
the interaction picture of $H_{\text{drive}}$, the tilted-frame operators evolve
as $\tilde{S}_k^\pm(t) = \tilde{S}_k^\pm e^{\pm i\Omega_k t}$. Rewriting the
bilinear spin products in the tilted frame and dropping the double-quantum terms
oscillating at $\pm 2\Omega$, the two transverse products yield identical
secular parts:
\begin{equation}
S_i^y S_j^y,\; S_i^z S_j^z \;\longrightarrow\; \frac{1}{4}(\tilde{S}_i^+
\tilde{S}_j^- + \tilde{S}_i^- \tilde{S}_j^+),
\label{eq:secular_identity}
\end{equation}
while $S_i^x S_j^x$ is time-independent and unchanged.

\subsubsection{Intra-group interactions}

Rewriting Eq.~\eqref{eq:intra} in Cartesian form using $S_i^+ S_j^- + S_i^-
S_j^+ = 2(S_i^x S_j^x + S_i^y S_j^y)$ and applying
Eq.~\eqref{eq:secular_identity} to the $S^y S^y$ and $S^z S^z$ terms yields:
\begin{equation}
\tilde{H}_{\text{intra}}^{ij} = \frac{J_{ij}}{8}(\tilde{S}_i^+ \tilde{S}_j^- +
\tilde{S}_i^- \tilde{S}_j^+) - \frac{J_{ij}}{2} S_i^x S_j^x.
\label{eq:intra_dressed}
\end{equation}
This retains the structure of Eq.~\eqref{eq:intra} but with the quantization
axis rotated from $z$ to $x$ and the coupling rescaled by a factor of $-1/2$:
the flip-flop strength becomes $J_{ij}/8$ and the Ising term $S_i^x S_j^x$ has
coefficient $-J_{ij}/2$.

\subsubsection{Inter-group NV-P1 interactions}

For the Ising Hamiltonian Eq.~\eqref{eq:inter}, applying
Eq.~\eqref{eq:secular_identity} converts $S_i^z S_j^z$ into flip-flop terms with
time dependence $e^{\pm i(\Omega_{\text{NV}} - \Omega_{\text{P1}})t}$. Under the
Hartmann-Hahn condition $\Omega_{\text{NV}} = \Omega_{\text{P1}} \equiv \Omega$,
these become time-independent:
\begin{equation}
\tilde{H}_{\text{inter}}^{ij} = \frac{J_{ij}}{4}(\tilde{S}_i^+ \tilde{S}_j^- +
\tilde{S}_i^- \tilde{S}_j^+).
\label{eq:inter_dressed}
\end{equation}
The Ising interaction that suppressed NV-P1 cross-relaxation is thus converted
into coherent spin exchange with flip-flop coupling $J_{ij}/4$, identical in
form to the lab-frame intra-group flip-flop interaction Eq.~\eqref{eq:intra}.
This coupling determines the transition rates in the spin diffusion model
(Appendix~\ref{sec:simulation}).

\begin{figure}
\includegraphics{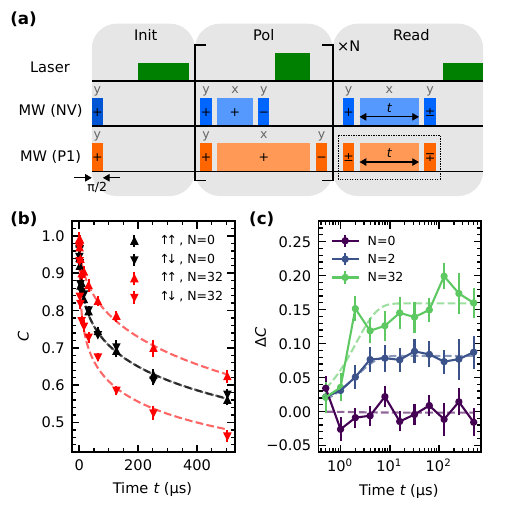}
\caption{Polarization transfer pulse sequences and polarization dynamics. (a)
Detailed pulse sequences. Light (dark) colors represent microwave pulses along
the $x$ ($y$) axis. The polarization transfer stage involves 5~\textmu{}s of DSL
under the HH condition and 5~\textmu{}s of laser illumination. In the subsequent
readout stage, the signal is obtained by sweeping the duration $t$ of the second
DSL sequence. For the NV (blue), the final $\pi/2$ pulse employs phase cycling
($\pm$) for common-mode noise rejection. For the P1 (orange), the pulses in the
dotted box prepare the P1 bath polarization parallel ($\uparrow\uparrow$) or
antiparallel ($\uparrow\downarrow$) to the NV. In the absence of these pulses,
the sequence measures the reference bare NV SL. (b) Normalized SL signal $C$ for
parallel and antiparallel configurations ($\Omega=6.40$~MHz). Data are shown for
the thermal state ($N=0$, black) and after iterative transfer ($N=32$, red).
Dashed lines indicate stretched exponential fits. (c) Signal difference $\Delta
C$ versus SL duration $t$ for varying cycle numbers $N$. Dashed lines are
exponential fits. Error bars represent s.e.m. \label{fig:2}}
\end{figure}

\section{Iterative polarization transfer and differential readout}

We probe the accumulation of P1 polarization using the pulse sequence depicted
in Fig.~\ref{fig:2}(a). To eliminate readout artifacts and isolate the signal
originating from the probed NV centers, we employ a common-mode rejection scheme
by subtracting two signals acquired with $\pi/2$ pulses of opposite
phase~\cite{kucsko_critical_2018}. Additionally, we normalize the DSL signal to
the bare NV SL decay measured in the absence of P1 driving. This allows us to
extract the P1 polarization dynamics mediated by dipolar interactions,
independent of intrinsic NV relaxation.

Fig.~\ref{fig:2}(b) shows the normalized SL signals $C$ obtained after $N=32$
transfer cycles for two different polarization configurations. We observe a
prominent asymmetry between parallel
($\uparrow_{\mathrm{NV}}\uparrow_{\mathrm{P1}}$) and antiparallel
($\uparrow_{\mathrm{NV}}\downarrow_{\mathrm{P1}}$) P1 bath configurations
relative to the NV spin. The parallel configuration suppresses energy-conserving
flip-flops, extending the DSL lifetime, whereas the antiparallel configuration
accelerates the decay. Thus, the difference signal $\Delta C =
C_{\uparrow\uparrow} - C_{\uparrow\downarrow}$ serves as a direct measure of the
net polarization in the resonantly addressed P1 ensemble.

The temporal evolution of $\Delta C$ [Fig.~\ref{fig:2}(c)] reveals the
polarization exchange dynamics during the readout DSL time $t$. The signals rise
and level off, indicating that the NV and P1 ensembles reach a transient
quasi-equilibrium. The dynamics are well described by an exponential function
$\Delta C(t) = A(1 - e^{-t/\tau_{\mathrm{eq}}})$ with a characteristic time
$\tau_{\mathrm{eq}} = 2.2 \pm 0.6$~\textmu{}s. This timescale is over two orders
of magnitude shorter than the NV SL relaxation time $T_{1\rho}^{\mathrm{NV}}
\approx 1.3$~ms. The distinct separation of timescales supports the
interpretation that the observed equilibration is driven by resonant NV-P1 spin
exchange, rather than by intrinsic NV relaxation channels.

We estimate the absolute polarization of the resonantly addressed P1 subgroup by
invoking conservation of total angular momentum in the rotating frame. From the
concentration ratio $n_{\text{P1}}/n_{\text{NV}} \approx 2.6$ and the
quasi-equilibrium signal amplitude $A$, we determine the P1 polarization as
$P^{\text{P1}} = P_0^{\text{NV}} (A/2) (1 + n_{\text{NV}}/n_{\text{P1}})$
(Appendix~\ref{sec:polarization}). Assuming an initial NV polarization of
$P_0^{\text{NV}} \approx 75\%$~\cite{robledo_spin_2011} and the saturation
amplitude $A_\infty \approx 0.143$ in the strong-driving limit [cf.
Fig.~\ref{fig:4}(b)], we estimate a P1 polarization of $P^{\text{P1}} \approx
7.4\%$. This corresponds to a polarization enhancement of approximately 740-fold
compared to the thermal equilibrium polarization of $\sim 0.010\%$ at room
temperature under the bias field. Converting this polarization into an effective
spin temperature via $T_{\text{spin}} = \hbar \gamma_e B / (2 k_B \tanh^{-1}
P^{\text{P1}})$ yields $T_{\text{spin}} \approx 405\ \mathrm{mK}$. These results
show that our protocol effectively prepares a low-entropy mesoscopic spin state
from a thermal mixed state.

\begin{figure}
\includegraphics{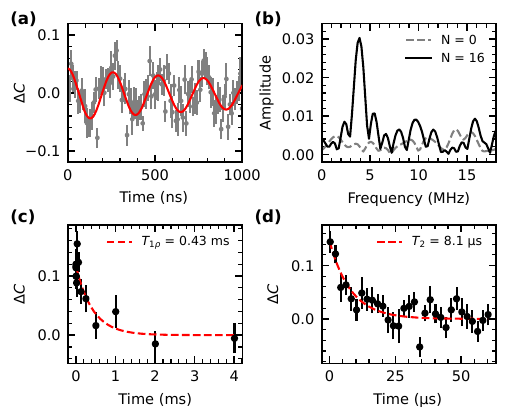}
\caption{Coherent control and relaxation dynamics of the polarized P1 ensemble.
(a) Collective Rabi oscillations of the P1 bath measured after $N=16$ transfer
cycles. The solid red line represents a fit to a damped cosine function. (b) FFT
spectra of the Rabi signal. A distinct frequency component appears for the
polarized state ($N=16$, black), whereas no signature is visible for the thermal
state ($N=0$, gray). (c) Rotating-frame spin relaxation time ($T_{1\rho}$)
measured via SL. (d) Measured Hahn-echo spin dephasing time ($T_2$). Dashed red
lines in (c) and (d) are fits to exponential decays. Error bars represent s.e.m.
\label{fig:3}}
\end{figure}

\section{Coherent control and relaxation of the polarized ensemble}
\label{sec:coherent}

\begin{figure}
\includegraphics{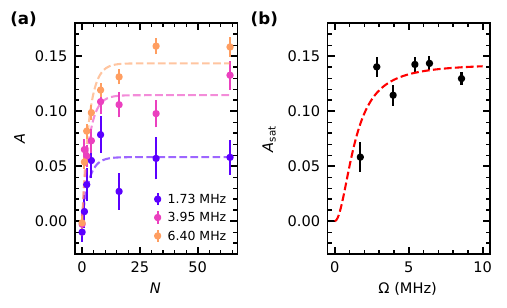}
\caption{Competition between coherent driving and disorder. (a)
Quasi-equilibrium amplitude $A$ versus cycle number $N$ for various P1 Rabi
frequencies $\Omega$. The signal saturates as the polarization transfer rate
balances with dissipation. Dashed lines are exponential fits. (b) Dependence of
the saturation amplitude $A_{\text{sat}}$ on $\Omega$. The signal enhancement at
higher Rabi frequencies indicates that strong driving suppresses the local
disorder, thereby facilitating spin transport across the ensemble. The red
dashed line is a fit to the phenomenological model described in the text. Error
bars indicate fitting uncertainties. \label{fig:4}}
\end{figure}

To verify the coherent nature of the polarized bath, we perform collective
manipulation. The readout stage likewise employs the HH condition to coherently
map the P1 ensemble spin polarization back onto the NV centers for optical
detection. By inserting a control pulse sequence between the polarization and
readout blocks, we observe clear Rabi oscillations of the P1 ensemble
[Fig.~\ref{fig:3}(a)]. These measurements are performed at $N=16$ cycles, well
into the saturation regime [cf. Fig.~\ref{fig:4}(a)], ensuring that the P1 bath
has reached its steady-state polarization. The fast Fourier transform (FFT) of
the signal shows a distinct frequency component associated with the driven P1
spins, which is absent in the unpolarized thermal state ($N=0$)
[Fig.~\ref{fig:3}(b)].

It is essential to distinguish this observation from conventional DEER
oscillations measured in a thermal bath~\cite{de_lange_controlling_2012}. In a
standard DEER measurement, the oscillating signal originates from modulation of
NV decoherence by fluctuating local fields of an unpolarized P1 bath. Such
measurements probe correlations, but do not indicate collective coherence of the
bath. In contrast, our measurement directly probes the population dynamics of
the P1 ensemble. The absence of oscillations at $N=0$ and their emergence at
$N=16$ confirm that we coherently manipulate a net mesoscopic polarization. The
oscillation amplitude grows with the number of polarization cycles and
saturates, and the oscillation frequency scales linearly with the microwave
drive amplitude (Appendix~\ref{sec:rabi_validation}).

We then characterize the relaxation dynamics of the spin-initialized P1
ensemble. The rotating-frame relaxation time, which sets the effective lifetime
for SL operations, is measured to be $T_{1\rho}^{\text{P1}} = 0.43 \pm 0.16$~ms
[Fig.~\ref{fig:3}(c)]. Notably, the measured $T_{1\rho}^{\text{P1}}$ is
comparable to the independently measured $T_1^{\text{P1}} \approx 0.67$~ms,
which indicates that rotating-frame relaxation is governed by the same intrinsic
processes that limit $T_1$, rather than by rapid polarization diffusion into the
surrounding unpolarized bulk. Furthermore, the Hahn-echo dephasing time is
measured to be $T_2^{\text{P1}} = 8.1 \pm 1.6$~\textmu{}s [Fig.~\ref{fig:3}(d)].
The ratio of this value to the NV dephasing time ($T_2^{\text{NV}} = 17.2 \pm
0.2$~\textmu{}s), $T_2^{\text{NV}} / T_2^{\text{P1}} \approx 2.1$, is of the
same order as the defect density ratio. This correspondence supports a picture
in which Hahn-echo dephasing in both ensembles is dominated by intragroup
dipolar flip-flop interactions, while intergroup Ising couplings are refocused
by the echo~\cite{bauch_decoherence_2020}.

\section{Saturation and disorder}
\label{sec:saturation}

To investigate the limits of the iterative polarization protocol, we monitor the
buildup of the polarization as a function of the cycle number $N$
[Fig.~\ref{fig:4}(a)]. The saturation indicates that the local P1 bath reaches a
steady state, in which polarization injection from the NV centers is
counterbalanced by dissipation. The buildup dynamics are well described by $A(N)
= A_{\text{sat}}(1 - e^{-N/N_{\text{sat}}})$, with characteristic cycle number
$N_{\text{sat}} \approx 3$. Relative to a single transfer cycle, 32 cycles
increase the quasi-equilibrium amplitude by a factor of $2.96 \pm 0.38$ at
$\Omega = 6.40$~MHz. Given a period of 10~\textmu{}s per cycle, this corresponds
to a saturation timescale of $\tau_{\text{sat}} \approx 30$~\textmu{}s, an order
of magnitude shorter than the $T_{1\rho}^{\text{P1}}$ measured in the dark. To
understand this rapid saturation, we employ a semiclassical master equation on a
disordered spin network (Appendix~\ref{sec:simulation}). Fitting the observed
saturation cycles yields an effective P1 relaxation time of
$T_{1\rho,\text{laser}}^{\text{P1}} \approx 31$~\textmu{}s during laser
illumination. This enhanced relaxation is consistent with laser-induced charge
dynamics under 532~nm illumination, including P1 photo-ionization and NV
charge-state
conversion~\cite{aslam_photo-induced_2013,giri_coupled_2018,%
marcks_quantum_2024,wang_manipulating_2023}, and is supported by the
power-dependent relaxation measurements (Appendix~\ref{sec:laser_depol}). The
model also reproduces the absolute polarization: the simulated saturation value
$P_\infty \approx 7.6\%$ agrees with the estimated $P^{\text{P1}} \approx 7.4\%$
within the uncertainty of the estimate.

Fig.~\ref{fig:4}(b) shows the dependence of the saturation amplitude
$A_{\text{sat}}$ on the Rabi frequency $\Omega$, reflecting the competition
between coherent driving strength and local disorder. In the hybrid electron
spin network, the random spatial distribution of P1 centers creates an
inhomogeneous energy landscape with characteristic disorder scale $W$. In the
weak driving regime ($\Omega < W$), energy mismatches between neighboring spins
suppress resonant flip-flop interactions, whereas in the strong driving regime
($\Omega > W$), these mismatches are overcome. This crossover is captured by
$A_{\text{sat}}(\Omega) = A_\infty \Omega^2/(\Omega^2 + W^2)$, and fitting
yields $W = 1.36 \pm 0.19$~MHz, interpreted as a local dipolar-detuning scale. A
Monte Carlo calculation for the same finite spin-network geometry gives
$W_{\mathrm{MC}} = 1.25 \pm 0.13$~MHz, supporting this MHz-scale interpretation.

The spatial extent of the steady-state polarization is governed by spin
diffusion within the network. Spin-network simulations give a root-mean-square
(rms) diffusion length $L_D \approx 6$~nm at the saturation timescale
$\tau_{\text{sat}} \approx 30~\mu$s (Appendix~\ref{sec:diffusion}). Although the
mean nearest-neighbor distance $d_{\text{NN}} \approx 14.8$~nm of the addressed
NV subgroup ($[\text{NV}] \approx 0.3$~ppm) exceeds this value, Poisson
statistics indicate that approximately 28\% of NV centers have at least one
neighbor within twice the rms displacement. The resulting polarization clusters,
where overlapping diffusion fronts reduce the local polarization gradient,
suppress outward spin diffusion and stabilize the steady-state polarization
lifetime.

\section{Conclusions and outlook}

In conclusion, we demonstrate that an iterative Hartmann-Hahn protocol with a
dense NV network generates a coherent mesoscopic spin state in a disordered P1
ensemble, achieving a 740-fold polarization enhancement over room-temperature
thermal equilibrium. The emergence of collective Rabi oscillations verifies the
coherent nature of the polarized state, while subsequent characterization
reveals long-lived spin-lock and Hahn-echo lifetimes. Rabi-frequency-dependent
saturation exposes a competition between coherent driving and local disorder.
The formation of local polarization clusters suppresses outward spin diffusion,
extending the polarization lifetime. The preparation, readout, and coherent
control demonstrated here, together with the quantitative limits on polarization
buildup, provide the basis for employing polarized dark electron spin ensembles
in sensing and simulation protocols.

Further optimization of the polarization process relies on a deeper
understanding and control of the underlying dissipation mechanisms, particularly
the charge dynamics identified during laser illumination. Future efforts in
sample synthesis will focus on balancing the NV-to-P1 ratio to optimize the
trade-off between polarization transfer efficiency and the total volume of the
available quantum resource. Beyond materials engineering, the implementation of
advanced control techniques such as Hamiltonian
engineering~\cite{schwartz_robust_2018, choi_robust_2020} and algorithmic
cooling~\cite{baugh_experimental_2005,zaiser_cyclic_2021} offers a pathway to
further reduce the effective spin temperature, by suppressing unwanted
depolarization and more efficiently extracting entropy from the spin ensemble.

The ability to prepare and manipulate these mesoscopic states opens broad
avenues for quantum sensing and simulation. In sensing, highly polarized P1
ensembles can serve as environment-assisted
sensors~\cite{goldstein_environment-assisted_2011, goldblatt_sensing_2024} or as
polarization sources for nuclear spins~\cite{pagliero_multispin-assisted_2018},
facilitating the generation of metrological states with enhanced
sensitivity~\cite{zheng_preparation_2022, wu_spin_2025-1}. As an analog quantum
simulator, this platform provides a controllable spin ensemble with tunable
dissipation to investigate non-equilibrium quantum thermodynamics. These
capabilities further enable the exploration of complex many-body phenomena in
disordered dipolar spin networks~\cite{choi_observation_2017,
selco_emergent_2025}, bridging the gap between controlled few-body quantum
systems and macroscopic spin ensembles.

\begin{acknowledgments}
T.Y. and H.C. acknowledge support from the National Research Foundation of Korea
(NRF) through the government of Korea (Grant Nos. 2021R1A2C3005905,
RS-2024-00466612, RS-2024-00413957, RS-2023-00258359), the Institute for Basic
Science (IBS) in Korea (Grant No. IBS-R034-D1), and Samsung Electronics Co., Ltd
(Grant No. IO260323-16108-01). J.L. acknowledges support from the NRF of Korea
(Grant Nos. RS-2025-25454922, RS-2024-00442710), IITP of Korea (Grant Nos.
RS-2025-02219034, RS-2025-25464252), and the KIST institutional program (Grant
No. 26E0011).
\end{acknowledgments}

\appendix

\section{Estimation of NV and P1 concentrations}
\label{sec:concentration}

We estimate the absolute densities of NV and P1 centers by comparing
experimental double electron-electron resonance (DEER) measurements with
numerical quantum dynamics simulations based on the Hamiltonians derived in
Sec.~\ref{sec:hamiltonian}. While it is well-established that the spin dephasing
rate $1/T_2$ scales linearly with the density of the surrounding magnetic spin
bath, determining the absolute concentration remains challenging because the
proportionality constant depends on the specific spin species, their spatial
distribution, and the experimental pulse
sequences~\cite{barry_sensitivity_2020}.

\begin{figure*}
\centering
\includegraphics{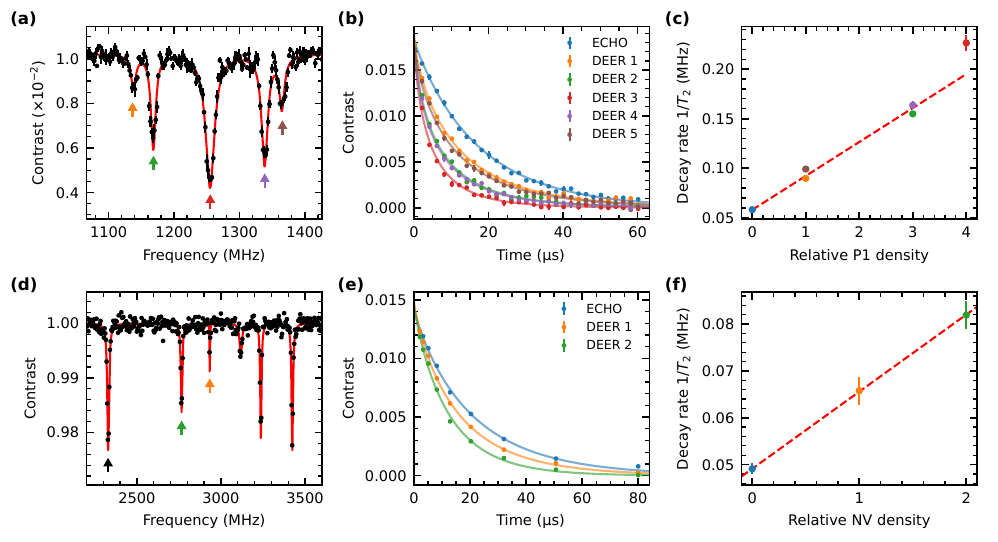}
\caption{Experimental characterization of P1 and NV spin baths.
(a) DEER spectrum of P1 centers. Colored arrows indicate the addressed resonance
frequencies corresponding to different hyperfine and Jahn-Teller subgroups.
(b) DEER decay traces for each P1 subgroup.
(c) Extracted dephasing rates ($1/T_2$) versus relative P1 density. Point colors
correspond to the resonance dips marked in (a).
(d) CW-ODMR spectrum of NV centers under a magnetic field aligned along $[1, 1,
1.5]$. The black arrow indicates the sensor NV resonance, and colored arrows
indicate the bath NV resonances.
(e) DEER decay traces for each NV group configuration ($\{1:0, 1:1, 1:2\}$).
(f) Extracted dephasing rates versus relative NV density. Point colors
correspond to the arrows in (d).
\label{fig:deer_exp}
}
\end{figure*}

\subsection{Experimental DEER measurements}

The experimental DEER spectra and decay traces are summarized in
Fig.~\ref{fig:deer_exp}. We measure the dephasing rates as a function of the
relative spin density for each subgroup. The DEER sequence applies a $\pi$ pulse
on the bath spins during a Hahn-echo sequence on the sensor spin, allowing
isolation of the dephasing contribution from the addressed bath spins.

For NV-P1 DEER, we address individual P1 hyperfine subgroups at the main
experimental field of 446~G. The P1 spectrum consists of five resolvable dips
arising from the $^{14}$N hyperfine interaction and Jahn-Teller distortion, with
population fractions $\{1,3,4,3,1\}/12$. By sequentially addressing different
subgroups, we measure the dephasing rate as a function of the relative P1
density.

For NV-NV DEER measurements, a magnetic field of approximately 200~G is aligned
along the $[1, 1, 1.5]$ crystallographic direction. This alignment ensures that
two of the four NV axis groups are energetically degenerate, allowing us to tune
the effective number of bath spins by addressing these overlapping resonances.
We perform measurements for three distinct sensor-to-bath configurations:
$\{1:0, 1:1, 1:2\}$, where the $1:0$ case corresponds to a standard Hahn-echo
(no bath spins flipped) and the $1:1$ and $1:2$ cases involve addressing one or
two bath groups, respectively. By performing a linear fit to the decay rates
extracted from these traces, we obtain the experimental slope
$\gamma_{\text{exp}}$ in units of MHz/group.

\subsection{Quantum dynamics simulation}

\begin{figure*}
\centering
\includegraphics{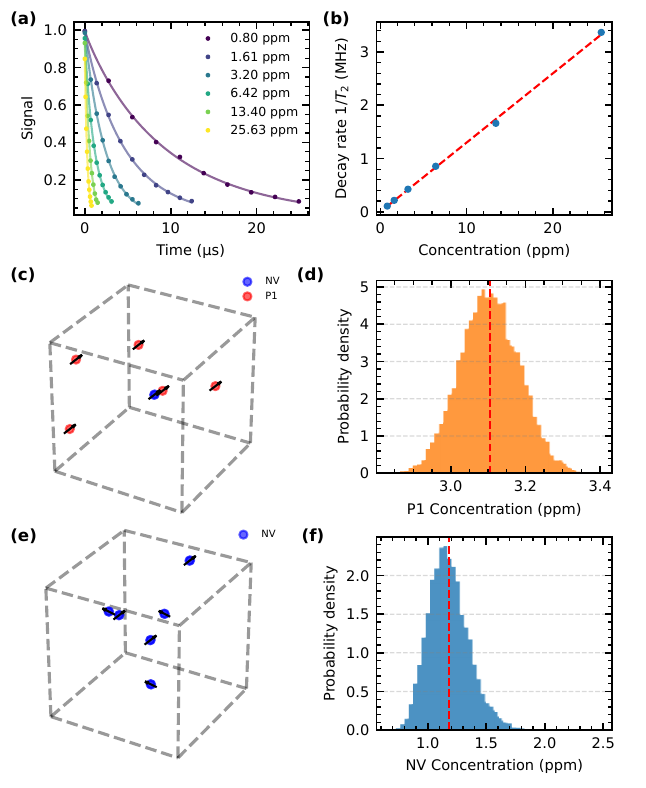}
\caption{Numerical simulation of DEER dynamics in disordered spin clusters.
(a) Simulated NV-P1 DEER decay traces at various P1 concentrations
(0.8--26~ppm).
(b) Simulated dephasing rates as a function of absolute P1 density. The dashed
line is a linear fit.
(c) Representative NV-P1 cluster configuration used in the simulation.
(d) Probability distribution of the estimated P1 concentration from Monte Carlo
uncertainty propagation.
(e) Representative NV cluster configuration used for NV-NV DEER simulations.
(f) Probability distribution of the estimated NV concentration from Monte Carlo
uncertainty propagation.
All dynamical results are averaged over 1000 random realizations.
\label{fig:deer_sim}
}
\end{figure*}

To calibrate the experimental slopes and extract absolute concentrations, we
perform numerical simulations [Fig.~\ref{fig:deer_sim}] using a disordered
dipolar spin model implemented in the QuTiP framework~\cite{lambert_qutip_2026}.
Spin positions are generated on a diamond lattice (lattice constant $a =
0.3567$~nm) within a cubic simulation volume. Each defect is assigned one of the
four $\langle 111 \rangle$ crystallographic orientations according to the
specified axis distribution. Configurations with nearest-neighbor distances less
than 1~nm are excluded.

For NV-P1 DEER, we model a cluster consisting of 1 NV sensor and 5 P1 bath spins
treated quantum mechanically, while for NV-NV DEER, we consider a cluster of 6
NV centers distributed among the relevant axis groups (e.g., $\{2, 2, 2, 0\}$).
The system Hamiltonian is constructed from Eqs.~\eqref{eq:intra} and
\eqref{eq:inter}: spins sharing the same type and axis interact via the full
secular Hamiltonian (flip-flop + Ising), while heterogeneous pairs (different
species or axes) interact via the Ising term only. For NV centers, the effective
spin-1/2 operators are scaled by $\sqrt{2}$ to account for the $|0\rangle
\leftrightarrow |{-}1\rangle$ subspace transitions in spin 1 matrix elements.

The decay traces are fitted to a stretched exponential function $S(t) = A
\exp[-(t/T_2)^\beta]$, from which the dephasing rate $1/T_2$ is extracted. To
capture ensemble-averaged behavior arising from intrinsic positional disorder,
we average the dynamics over 1000 independent random spatial realizations for
each density point. The concentration-dependent simulation is performed at
multiple target densities spanning 0.8--200~ppm, with the evolution time scaled
inversely with concentration to ensure adequate sampling of the decay dynamics.

\subsection{Concentration determination}

The simulated dephasing rates are fitted to a linear model $\Gamma = \alpha
\cdot n$, yielding the numerical dephasing coefficient $\alpha$ [MHz/ppm] for
each subgroup configuration. To account for the sensor-bath configurations used
in the experiment, we calculate the simulated DEER slope $K$
[MHz/(group$\cdot$ppm)] by evaluating the expected dephasing rate for each group
configuration at a total density of 1~ppm.

The total ensemble density is determined by the ratio $n_{\text{total}} =
\gamma_{\text{exp}} / K$. To rigorously propagate the uncertainties from both
the experimental fit and the numerical simulation, we employ a Monte Carlo
analysis ($N=10{,}000$). In each iteration, values for the experimental slopes
and numerical dephasing coefficients are sampled from their respective Gaussian
distributions.

Following this estimation procedure, we determine the total NV density to be
$[\text{NV}] = 1.2 \pm 0.2$~ppm and the total P1 density to be $[\text{P1}] =
3.1 \pm 0.1$~ppm. The corresponding concentration ratio is
$n_{\text{P1}}/n_{\text{NV}} \approx 2.6$. These values are consistent with
independent estimates obtained by applying established dephasing-rate
calibrations~\cite{bauch_decoherence_2020,wang_manipulating_2023,%
chernyavskiy_double_2025} to the measured $T_2^* = 1.58 \pm 0.01$~\textmu{}s and
$T_2 = 17.2 \pm 0.2$~\textmu{}s of our sample.

\section{Estimation of P1 polarization}
\label{sec:polarization}

We derive the relationship between the measured signal contrast and the P1 bath
polarization by considering conservation of rotating-frame polarization during
double spin-locking at the Hartmann-Hahn condition. Let $n_{\text{NV}}$ and
$n_{\text{P1}}$ denote the spin densities of the addressed NV and P1 subgroups,
respectively, and let $P_{\text{NV},0}$ be the initial NV polarization after
optical pumping. When the readout signal reaches a plateau under the
Hartmann-Hahn condition, the total rotating-frame polarization is approximately
conserved and redistributed between the two spin species in an ensemble-averaged
sense.

For the parallel ($\uparrow\uparrow$) configuration, where the P1 bath
polarization is aligned with the NV polarization, the total initial polarization
is $P_{\text{tot}} = n_{\text{NV}}P_{\text{NV},0} + n_{\text{P1}}P_{\text{P1}}$.
At the readout plateau, the optically probed NV ensemble and the coupled P1
ensemble are described by a common ensemble-averaged polarization
$P_{\uparrow\uparrow}$:
\begin{equation}
n_{\text{NV}}P_{\text{NV},0} + n_{\text{P1}}P_{\text{P1}} =
(n_{\text{NV}}+n_{\text{P1}})P_{\uparrow\uparrow}.
\label{eq:cons_parallel}
\end{equation}
For the antiparallel ($\uparrow\downarrow$) configuration, the P1 polarization
contributes with opposite sign. The readout plateau is described by a common
ensemble-averaged polarization $P_{\uparrow\downarrow}$:
\begin{equation}
n_{\text{NV}}P_{\text{NV},0} - n_{\text{P1}}P_{\text{P1}} =
(n_{\text{NV}}+n_{\text{P1}})P_{\uparrow\downarrow}.
\label{eq:cons_antiparallel}
\end{equation}

The spin-lock signal used in the polarization estimate is the normalized
phase-cycled differential PL signal. For each readout point, we acquire two
phase-cycled PL signals after the final $\pm x$ mapping pulses and define
\begin{equation}
\delta I(t) = I_-(t) - I_+(t).
\end{equation}
The normalized signal is
\begin{equation}
C(t) = \frac{\delta I_{\mathrm{NV-P1}}(t)}{\delta I_{\mathrm{NV}}(t)},
\end{equation}
where $\delta I_{\mathrm{NV-P1}}(t)$ is the differential PL signal measured in
the NV-P1 spin-lock readout and $\delta I_{\mathrm{NV}}(t)$ is the corresponding
bare NV reference measured under the same readout condition. Under the leading
linear readout calibration, $C_s = P_s / P_{\mathrm{NV},0}$ for $s =
\uparrow\uparrow, \uparrow\downarrow$, so that
\begin{equation}
P_{\uparrow\uparrow} = P_{\mathrm{NV},0} C_{\uparrow\uparrow}, \qquad
P_{\uparrow\downarrow} = P_{\mathrm{NV},0} C_{\uparrow\downarrow}.
\end{equation}
Subtracting Eq.~\eqref{eq:cons_antiparallel} from Eq.~\eqref{eq:cons_parallel}
and substituting the calibration relations yields
\begin{equation}
P_{\text{P1}} = P_{\text{NV},0} \frac{\Delta C}{2} \left(1 +
\frac{n_{\text{NV}}}{n_{\text{P1}}}\right),
\label{eq:p1_polarization}
\end{equation}
where $\Delta C = C_{\uparrow\uparrow} - C_{\uparrow\downarrow}$. In the
experiment, we use the plateau value $A$ of $\Delta C(t)$ as the
ensemble-averaged quasi-equilibrium readout amplitude.

We also estimate the effect of residual nonlinear readout response. With the
sign convention above, the phase-cycled PL signals expand as $I_\pm(P) =
\alpha_0 \mp \alpha_1 P + \alpha_2 P^2 \mp \alpha_3 P^3 + \cdots$, so the
differential signal is
\begin{equation}
\delta I(P)=I_-(P)-I_+(P)=2\left(\alpha_1 P+\alpha_3 P^3+\cdots\right).
\end{equation}
All even-order nonlinearities cancel by symmetry, so the leading residual
nonlinearity is cubic. Writing $\eta=\alpha_3/\alpha_1$, the normalized plateau
signal is
\begin{equation}
C_s = \frac{P_s+\eta P_s^3}{P_{\mathrm{NV},0}+\eta P_{\mathrm{NV},0}^3}+\cdots.
\end{equation}
Expanding to first order in $\eta$ gives
\begin{equation}
\frac{\delta P^{\mathrm{P1}}}{P^{\mathrm{P1}}} = \eta
P_{\mathrm{NV},0}^2\left[1-\left(C_{\uparrow\uparrow}^2+%
C_{\uparrow\uparrow}C_{\uparrow\downarrow}+%
C_{\uparrow\downarrow}^2\right)\right]+\cdots.
\end{equation}
Using the conservative bound
$|C_{\uparrow\uparrow}|,|C_{\uparrow\downarrow}|\le1$ gives
\begin{equation}
\left|\frac{\delta P^{\mathrm{P1}}}{P^{\mathrm{P1}}}\right| \lesssim
2|\eta|P_{\mathrm{NV},0}^2.
\end{equation}
Even for an illustrative 1\% cubic nonlinearity, $|\eta|\sim0.01$, this gives a
relative correction $\lesssim1.1\%$ for $P_{\mathrm{NV},0}\simeq0.75$, smaller
than the uncertainty from the defect-density ratio and the fitted plateau
amplitude.
We note that this quasi-equilibrium is ensemble-averaged: because local dipolar
couplings are spatially inhomogeneous, $P_{\uparrow\uparrow}$ and
$P_{\uparrow\downarrow}$ denote ensemble-averaged plateau polarizations sampled
by the optical NV readout, not uniform site-by-site values.

The validity of using the plateau amplitude is supported by the experimentally
observed separation of timescales. The differential spin-lock signal reaches a
plateau according to $\Delta C(t)=A(1-e^{-t/\tau_{\mathrm{eq}}})$, with
$\tau_{\mathrm{eq}}=2.2\pm0.6~\mu$s. This exchange time is much shorter than the
relaxation times that do not conserve rotating-frame polarization,
$T_{1\rho}^{\mathrm{NV}}\approx1.3$~ms and
$T_{1\rho}^{\mathrm{P1}}\approx0.43$~ms. Relaxation loss during the formation of
the ensemble-averaged quasi-equilibrium plateau is therefore negligible.

We also consider possible leakage to unaddressed P1 spectral groups. The density
$n_{\text{P1}}$ in Eq.~\eqref{eq:p1_polarization} denotes the resonantly
addressed P1 spectral subgroup, not the total P1 density. The addressed P1
resonance has population fraction $3/12$, while the addressed NV transition
corresponds to one NV crystallographic orientation. Since both addressed
fractions are one quarter of the corresponding total ensembles, the ratio
$n_{\text{P1}}/n_{\text{NV}}$ is equal to the measured total concentration
ratio.

Leakage to unaddressed P1 spectral groups would require spin flip-flop exchange
between different hyperfine or Jahn-Teller configurations. This process is
strongly suppressed under the spin-lock condition. The microwave carrier
resonantly drives the targeted P1 spectral group, while the other resolved P1
spectral groups are separated in frequency by the $^{14}\mathrm{N}$ hyperfine
interaction and the Jahn-Teller orientations. Under continuous spin locking,
flip-flop terms involving these unaddressed groups oscillate rapidly in the
rotating frame and become non-secular when the spectral mismatch is larger than
the relevant dipolar exchange rate. The remaining secular couplings from
unaddressed spins mainly contribute to the static dipolar disorder quantified by
$W$, rather than producing efficient polarization leakage.

Any leakage during the polarization-buildup cycles would act as an additional
loss channel for the resonantly addressed subgroup. It would reduce the
steady-state addressed-channel polarization before readout and would therefore
be already reflected in the measured saturation amplitude.
If a weak residual leakage channel remains during readout, its effect can be
represented phenomenologically as an attenuation of the addressed-channel
differential contrast,
\begin{equation}
\Delta C_{\mathrm{meas}} = (1-\eta_{\mathrm{leak}}) \Delta C^{(0)}, \qquad
0\le\eta_{\mathrm{leak}}<1,
\end{equation}
where $\Delta C^{(0)}$ is the differential contrast that would be measured from
the addressed subgroup in the absence of readout leakage, and
$\eta_{\mathrm{leak}}$ is the fractional leakage. Substitution into
Eq.~\eqref{eq:p1_polarization} gives
\begin{equation}
P_{\mathrm{P1}}^{\mathrm{calc}} = (1-\eta_{\mathrm{leak}})
P_{\mathrm{P1}}^{\mathrm{real}},
\end{equation}
where $P_{\mathrm{P1}}^{\mathrm{real}}$ denotes the polarization of the
resonantly addressed P1 subgroup present at the beginning of the readout.
Therefore, residual readout leakage to unaddressed P1 spectral groups would lead
to an underestimate of the actual addressed-subgroup polarization, not an
overestimate. It would not artificially enhance the inferred value.

\section{Numerical simulation of polarization dynamics}
\label{sec:simulation}

\subsection{Rate equation model}

Polarization transport in the spin-locked system is modeled by a semiclassical
master equation derived from Fermi's Golden
Rule~\cite{zu_emergent_2021,choi_depolarization_2017}. Both intra-group
(Eq.~\eqref{eq:intra_dressed}) and inter-group (Eq.~\eqref{eq:inter_dressed})
flip-flop interactions contribute to the transition rates, with dressed-frame
coupling strengths $J_{ij}/8$ and $J_{ij}/4$, respectively.

In the presence of disorder, each spin $i$ has a local detuning $\delta_i$ from
the drive frequency, so that the effective Rabi frequency becomes
$\Omega_{\text{eff},i} = \sqrt{\Omega^2 + \delta_i^2}$. The tilt angle of the
effective field away from the $z$ axis satisfies $\sin\theta_i = \Omega /
\Omega_{\text{eff},i}$ and $\cos\theta_i = \delta_i / \Omega_{\text{eff},i}$.
Only the components of the lab-frame operators transverse to the local effective
field participate in the dressed-frame flip-flop: the $S^zS^z$ term is projected
by $\sin\theta_i \sin\theta_j$, the $S^yS^y$ term by unity, and the $S^xS^x$
term by $\cos\theta_i \cos\theta_j$. With the lab-frame coefficients of
Eqs.~\eqref{eq:intra} and \eqref{eq:inter}, the effective coupling entering the
transition rate is
\begin{equation}
\tilde{J}_{ij} = \frac{J_{ij}}{4} \times \begin{cases}
\sin\theta_i \sin\theta_j - \tfrac{1}{2} - \tfrac{1}{2}\cos\theta_i \cos\theta_j
& \text{(intra-group)}, \\
\sin\theta_i \sin\theta_j & \text{(inter-group)},
\end{cases}
\end{equation}

By Fermi's Golden Rule, the transition rate for a flip-flop process between
spins $i$ and $j$ is given by the spectral overlap of the two spin lineshapes at
the effective detuning $\Delta_{ij}^{\text{eff}} = \Omega_{\text{eff},i} -
\Omega_{\text{eff},j}$:
\begin{equation}
R_{ij} = 2 |\tilde{J}_{ij}|^2 \frac{\Gamma}{\Gamma^2 +
(\Delta_{ij}^{\text{eff}})^2},
\end{equation}
where $\Gamma = 0.15$~MHz is the half-width at half-maximum of the Hartmann-Hahn
resonance [Fig.~\ref{fig:1}(d)].

In the continuum limit, polarization transport driven by flip-flop interactions
is described by the spin diffusion equation:
\begin{equation}
\frac{\partial P(\mathbf{r}, t)}{\partial t} = D \nabla^2 P(\mathbf{r}, t) -
\frac{P(\mathbf{r}, t)}{T_{1\rho}},
\end{equation}
where $D$ is the spin diffusion coefficient and $T_{1\rho}$ is the
rotating-frame relaxation time. For a disordered discrete spin network, we solve
the corresponding lattice master equation directly:
\begin{equation}
\frac{dP_i}{dt} = \sum_j R_{ij} (P_j - P_i) - \frac{P_i}{T_{1\rho, i}},
\end{equation}
where $R_{ij}$ is the pairwise polarization transfer rate, the first term
describes polarization exchange with all neighboring spins, and the second term
accounts for intrinsic spin-lattice relaxation at each site.

\subsection{Network model implementation}

To numerically simulate the polarization dynamics, we implement a discrete spin
network model consisting of randomly distributed NV and P1 centers. Spins are
placed at random positions, drawn uniformly, within a cubic simulation box of
side length $L$. Configurations with nearest-neighbor distances less than 1~nm
are excluded. The spin density is set to match the addressed P1 subgroup
concentration of $\sim$0.8~ppm ($= 3.1$~ppm $\times 3/12$), corresponding to a
mean inter-spin spacing of $d_{\text{avg}} = n^{-1/3} \approx 19.4$~nm, where $n
= c_{\text{ppm}} \times 1.76 \times 10^{-4}~\text{nm}^{-3}$ is the number
density.

For each pair of spins $(i, j)$, the dipolar coupling strength is computed as
$J_{ij} = J_0 (1 - 3\cos^2\theta_{ij}) / r_{ij}^3$, where $r_{ij}$ is the
inter-spin distance and $\theta_{ij}$ is the angle between the inter-spin vector
and the external magnetic field.

Local disorder arises from static longitudinal dipolar fields generated by the
surrounding electron-spin bath. In the transport model, we represent this effect
by a random detuning $\delta_i$ with standard deviation $W = 1.36$~MHz, where
$W$ is obtained from the saturation-amplitude fit in Sec.~\ref{sec:saturation}.

As a scale check, we independently estimated the static local detuning by Monte
Carlo sampling of random diamond-lattice configurations. For each target spin,
the local detuning is
\begin{equation}
\delta_i = \sum_{k \neq i} A_{ik}^{zz} m_k,
\end{equation}
where $m_k$ is the physical $S_k^z$ projection of bath spin $k$. We include all
P1 spins at the full bath density $[\mathrm{P1}] \simeq 3.1$~ppm, and the
resonantly addressed NV subgroup (with $m_{\mathrm{NV}} = 0$ and $-1$ with equal
probability). Unaddressed NV centers, initialized into $\lvert0\rangle$, carry
zero longitudinal projection and are excluded. This gives
\begin{equation}
W_{\mathrm{MC}} = \mathrm{std}(\delta_i) = 1.25 \pm 0.13~\mathrm{MHz},
\end{equation}
where the uncertainty is from bootstrap resampling. This agrees with the fitted
$W = 1.36 \pm 0.19$~MHz within combined uncertainties.
All reported results are averaged over 1000 independent random realizations of
spin positions to account for positional disorder.

\subsection{Spin diffusion and polarization transport}
\label{sec:diffusion}

\begin{figure}
\centering
\includegraphics{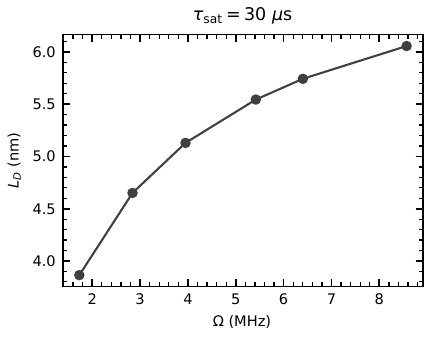}
\caption{Spin-network estimate of the polarization diffusion length.
Root-mean-square displacement $L_D = \sqrt{\text{MSD}(\tau_{\text{sat}})}$
evaluated at $\tau_{\text{sat}} = 30~\mu$s for the largest simulated system
($N_{\text{P1}} = 800$), plotted as a function of Rabi frequency $\Omega$. At
the experimental operating point $\Omega = 6.40$~MHz, the simulation gives $L_D
\approx 5.7$~nm.
\label{fig:diffusion}
}
\end{figure}

The MSD of the polarization distribution is defined as:
\begin{equation}
\text{MSD}(t) = \frac{\sum_i P_i(t) \, r_i^2}{\sum_i P_i(t)},
\end{equation}
where $r_i$ is the distance of spin $i$ from the source NV. Starting from a
single polarized NV center, we track the spin diffusion of polarization into the
surrounding P1 bath. Rather than extracting a continuum diffusion coefficient
from the disordered finite network, we characterize the spatial extent relevant
to the experiment by the root-mean-square (rms) diffusion length evaluated at
the saturation timescale $\tau_{\text{sat}} = N_{\text{sat}} \times
T_{\text{cycle}} \approx 30~\mu$s:
\begin{equation}
L_D \equiv \sqrt{\langle\Delta r^2(\tau_{\text{sat}})\rangle} =
\sqrt{\text{MSD}(\tau_{\text{sat}})}.
\end{equation}
Using the largest simulated system ($N_{\text{P1}} = 800$), we obtain $L_D
\approx 5.7$~nm at the experimental Rabi frequency $\Omega = 6.40$~MHz
(Fig.~\ref{fig:diffusion}).

The fraction of NV centers that form polarization clusters is estimated by
Poisson statistics. For a subgroup density $[\text{NV}]_\text{sub} = 1.2/4
\approx 0.30$~ppm (addressed axis), the mean number of neighboring NV centers
within a sphere of radius $2L_D$ is $\lambda = n_{\text{NV}} \times
\frac{4\pi}{3}(2L_D)^3$, giving a cluster probability $P_{\text{cluster}} = 1 -
e^{-\lambda} \approx 28\%$ at $\Omega = 6.40$~MHz.

\subsection{Iterative polarization transfer simulation}

\begin{figure}
\centering
\includegraphics{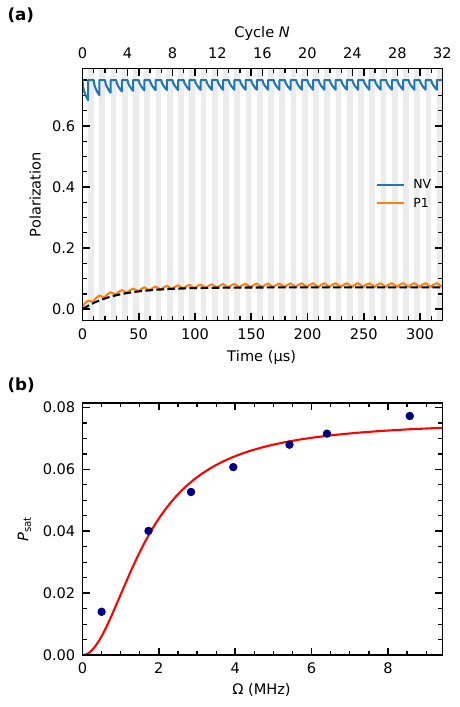}
\caption{Iterative polarization transfer simulation.
(a) Simulated NV (blue) and P1 (orange) polarization dynamics under the
iterative Hartmann-Hahn protocol at $\Omega = 6.40$~MHz. The NV polarization is
averaged over all NV centers and 1000 disorder realizations. The P1 polarization
is averaged over the 8 nearest P1 centers to the NV source. The black dashed
line is a saturation fit $A(N) = P_{\text{sat}}(1 - e^{-N/N_{\text{sat}}})$.
Gray shading indicates the laser initialization phase.
(b) Saturation polarization $P_{\text{sat}}$ versus Rabi frequency $\Omega$. The
red line is a fit to $P_{\text{sat}}(\Omega) = P_\infty \Omega^2 / (\Omega^2 +
\Omega_c^2)$. Error bars represent $1\sigma$ fitting uncertainties.
\label{fig:iterative_sim}
}
\end{figure}

To reproduce the experimentally observed saturation behavior, we incorporate a
two-phase model that accounts for the different relaxation dynamics during the
spin-lock (dark) and re-initialization (laser-on) phases of each polarization
cycle.

Each polarization transfer cycle consists of two phases: (1) a 5~\textmu{}s
interaction phase, during which NV and P1 spins interact under the Hartmann-Hahn
condition with P1 relaxation governed by $T_{1\rho}^{\text{P1}} =
430~\text{\textmu{}s}$, and (2) a 5~\textmu{}s NV initialization phase, during
which NV spins are optically repolarized to $P_{\text{NV},0} = 75\%$ while P1
spins experience enhanced relaxation with $T_{1\rho,\text{laser}}^{\text{P1}}
\approx 31~\text{\textmu{}s}$. The total cycle period is 10~\textmu{}s, and the
simulation propagates the polarization state for up to 32 cycles.

The significantly shortened effective relaxation time during laser illumination
is required to reproduce the observed saturation cycle number $N_{\text{sat}}
\approx 3$. Such rapid depolarization is consistent with photo-ionization of P1
centers and associated charge-state dynamics within the local spin network under
532~nm illumination.

The two-phase simulation (Fig.~\ref{fig:iterative_sim}) reproduces the
Rabi-frequency dependence of the saturation amplitude $A_{\text{sat}}(\Omega)$,
exhibiting the crossover behavior governed by the competition between driving
strength and disorder.
The model reproduces $N_{\text{sat}} \approx 3$, the Rabi-frequency dependence
of $A_{\text{sat}}(\Omega)$, and the absolute scale of the polarization: the
simulated saturation polarization $P_\infty \approx 7.6\%$ agrees with the
experimentally estimated value $P^{\text{P1}} \approx 7.4\%$ within the
uncertainty of the estimate, despite the neglect of quantum correlations among
strongly coupled spins and the use of a spatially uniform laser-induced
relaxation time.

We also tested the robustness of the extracted disorder scale to the chosen
crossover function. Fitting the same saturation-amplitude data to
$A_{\text{sat}} = A_\infty[1-\exp(-\Omega^2/2W^2)]$ gives $W = 1.33 \pm
0.11$~MHz with reduced $\chi^2 = 5.58$, compared with $W = 1.36 \pm 0.19$~MHz
and reduced $\chi^2 = 7.82$ for
$A_{\text{sat}}(\Omega)=A_\infty\Omega^2/(\Omega^2+W^2)$. The two values agree
within uncertainty, so the inferred MHz-scale disorder is not sensitive to the
exact two-parameter crossover form.

\subsection{Power-dependent P1 depolarization under laser illumination}
\label{sec:laser_depol}

As an independent check of the laser-induced depolarization used in the
two-phase model, we measured the P1 rotating-frame relaxation under 532~nm
illumination on a comparable dense diamond sample, DNV-B14 (Element Six,
$[\mathrm{NV}] = 4.5$~ppm, $[\mathrm{N}] = 8.5$~ppm). P1 polarization was first
prepared by the iterative Hartmann--Hahn protocol ($N=16$ cycles,
$\Omega=6.40$~MHz). We then applied a variable-duration 532~nm pulse while
maintaining the P1 spin-lock and read out the residual P1 polarization from the
differential spin-lock contrast.

\begin{figure}
\centering
\includegraphics{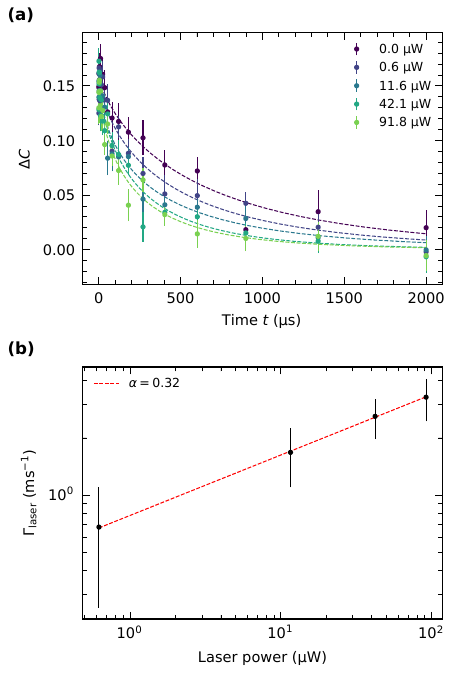}
\caption{Power-dependent P1 depolarization under laser illumination on DNV-B14.
(a) Differential spin-lock contrast $\Delta C$ versus spin-lock duration at five
532~nm laser powers, with stretched-exponential fits.
(b) Dark-subtracted depolarization rate $\Gamma_\mathrm{laser}$ versus laser
power on log-log axes. The dashed line is an empirical power-law fit with
exponent $\alpha = 0.32 \pm 0.12$.
\label{fig:laser_depol}}
\end{figure}

The decay at each laser power is well described by a stretched exponential,
\begin{equation}
S(\tau)=\exp[-(\tau/T)^\beta].
\end{equation}
The fitted timescale decreases from $629\pm101~\mu$s under near-dark conditions
to $206\pm33~\mu$s at 92~$\mu$W, with $\beta\approx0.62$--$0.76$
[Fig.~\ref{fig:laser_depol}(a)]. After subtracting the near-dark rate
$\Gamma_\mathrm{dark}=1.59\pm0.26~\mathrm{ms}^{-1}$, the laser-induced
depolarization rate follows an empirical sublinear power dependence
[Fig.~\ref{fig:laser_depol}(b)],
\begin{equation}
\Gamma_\mathrm{laser}(P_\mathrm{L})
=
\frac{1}{T(P_\mathrm{L})}
-
\Gamma_\mathrm{dark}
\propto
P_\mathrm{L}^{\alpha},
\qquad
\alpha = 0.32 \pm 0.12 .
\end{equation}
We interpret this exponent as an ensemble-averaged empirical trend rather than a
microscopic photon order. Possible charge-mediated channels include NV
charge-state cycling and direct P1 photo-ionization under 532~nm
illumination~\cite{aslam_photo-induced_2013,giri_coupled_2018,%
marcks_quantum_2024,wang_manipulating_2023}.

For comparison with the 5~$\mu$s laser re-initialization step in the
polarization-transfer sequence, we define an effective time constant by
\begin{equation}
\exp(-t_\mathrm{L}/T_\mathrm{eff})
=
\exp[-(t_\mathrm{L}/T)^\beta],
\qquad
t_\mathrm{L}=5~\mu\mathrm{s}.
\end{equation}
At laser powers of 11--92~$\mu$W, this gives
$T_\mathrm{eff}\approx58$--$72~\mu$s. This value is larger than the
simulation-inferred $T_{1\rho,\mathrm{laser}}^\mathrm{P1}\approx31~\mu$s for the
main-text sample, but is comparable in scale. The remaining difference is
consistent with sample-to-sample variation in defect density, optical intensity,
and local charge environment.

\section{Independent validation of collective Rabi oscillations}
\label{sec:rabi_validation}

To independently validate the collective Rabi interpretation of the oscillations
reported in Sec.~\ref{sec:coherent} [Figs.~\ref{fig:3}(a) and \ref{fig:3}(b)],
we performed additional measurements on a second diamond sample, DNV-B14. The
measurements used the same iterative Hartmann--Hahn protocol as the primary
sample, with readout via the differential spin-lock contrast $\Delta C$.

These measurements also distinguish driven ensemble Rabi oscillations from
intrinsic NV--P1 pair-exchange dynamics. For the addressed P1 subgroup density
in the primary sample, $[\mathrm{P1}]_\mathrm{addr}\simeq 0.775$~ppm, the mean
nearest-neighbor NV--P1 distance is $r_\mathrm{nn}\simeq 10.8$~nm, corresponding
to $J\sim J_0/r_\mathrm{nn}^3\simeq 42$~kHz. This characteristic scale is far
below the observed MHz-scale oscillations, and rare closer pairs would
contribute a broad $J\propto r^{-3}$ distribution rather than a narrow
drive-defined frequency.

\subsection{Cycle-number dependence of the oscillation amplitude}

\begin{figure}
\centering
\includegraphics{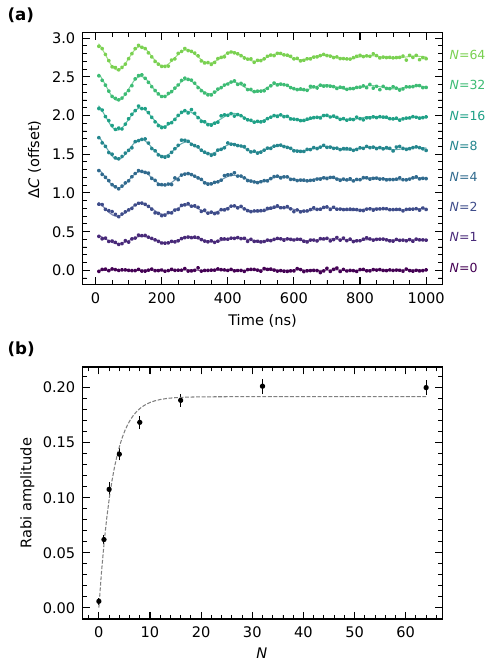}
\caption{Cycle-number dependence of P1 Rabi oscillations on DNV-B14.
(a)~Differential contrast $\Delta C(t)$ at $N = 0$--$64$ (offset for clarity).
No oscillations are visible at $N = 0$; they grow monotonically with $N$ at a
frequency of $7.0$--$7.2$~MHz.
(b)~Fitted Rabi amplitude vs.\ $N$ with exponential saturation fit $A(N) =
A_\mathrm{sat}(1 - e^{-N/N_\mathrm{sat}})$ (dashed line; $A_\mathrm{sat} = 0.192
\pm 0.004$, $N_\mathrm{sat} = 2.79 \pm 0.17$).
\label{fig:rabi_cycles}}
\end{figure}

Fig.~\ref{fig:rabi_cycles} shows the dependence of the Rabi signal on the number
of polarization cycles. At $N = 0$, the fitted amplitude is negligible ($|A| =
0.006$), consistent with the absence of net P1 polarization. For $N \geq 1$, the
amplitude increases monotonically and saturates with $N_\mathrm{sat} = 2.79 \pm
0.17$, close to the saturation scale of the polarization buildup in
Fig.~\ref{fig:4}(a). Thus, the oscillation amplitude tracks the accumulated P1
magnetization, while the oscillation frequency remains set by the microwave
drive.

\subsection{Drive-amplitude dependence of the oscillation frequency}

\begin{figure}
\centering
\includegraphics{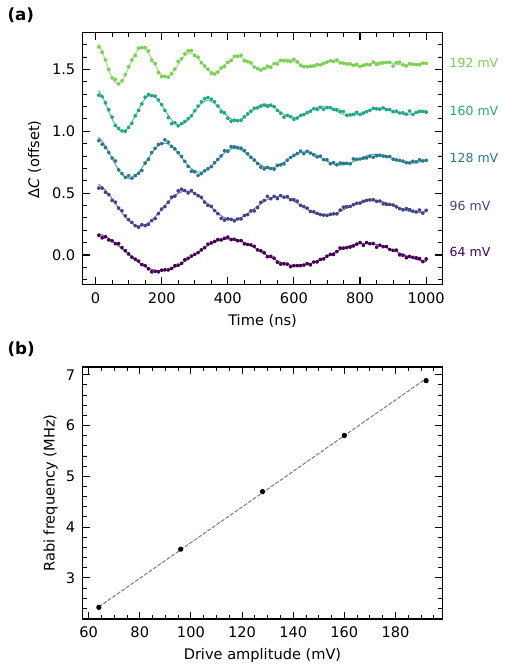}
\caption{Drive-amplitude dependence of P1 Rabi oscillations on DNV-B14.
(a)~Differential contrast $\Delta C(t)$ at five drive amplitudes (64--192~mV,
offset for clarity).
(b)~Extracted Rabi frequency vs.\ drive amplitude, with a linear fit (dashed
line; slope $0.0351 \pm 0.0001$~MHz/mV, intercept $0.184 \pm 0.011$~MHz).
\label{fig:rabi_drive}}
\end{figure}

Fig.~\ref{fig:rabi_drive} shows that the P1 Rabi frequency increases linearly
with microwave drive amplitude. A linear fit gives $f = (0.0351 \pm
0.0001)\,V_\mathrm{mV} + (0.184 \pm 0.011)$~MHz. This drive-defined scaling is
the expected behavior for driven Rabi oscillations and is inconsistent with an
intrinsic pair-exchange oscillation set by a static NV--P1 dipolar coupling.

\end{document}